\documentclass[twocolumn,showpacs,preprintnumbers,amsmath,amssymb,a4paper]{revtex4-1} 

\usepackage{amsmath,amssymb,bbm,epsfig,booktabs,natbib,bbold}
\usepackage[colorlinks=true,pdfstartview=FitH,linkcolor=blue,citecolor=blue,urlcolor=blue]{hyperref}
\usepackage[table]{xcolor} 
\newcommand{\be}{\begin{equation}}
\newcommand{\ee}{\end{equation}}
\newcommand{\ba}{\begin{array}}
\newcommand{\ea}{\end{array}}
\newcommand{\bqa}{\begin{eqnarray}}
\newcommand{\eqa}{\end{eqnarray}}

\newcommand{\um}{\mathbbm{1}}

\newcommand{\ket}[1]{\ensuremath{| #1 \rangle}}
\newcommand{\prj}[1]{\ensuremath{| #1 \rangle \langle #1 |}}

\newcommand{\matel}[3]{\ensuremath{\langle #1 | #2 | #3 \rangle}}

\newcommand{\ie}{{\it i.e.}}

\begin{document}

\title{Hierarchies of multi-partite entanglement}

\author{Federico Levi and Florian Mintert}

\affiliation{Freiburg Insitute for Advanced Studies, Albert-Ludwigs University of Freiburg, Albertstr. 19, 79104 Freiburg, Germany}

\date{\today}

\begin{abstract}
We derive hierarchies of separability criteria that identify the different degrees of entanglement ranging from bipartite to genuine multi-partite in mixed quantum states of arbitrary size. 
\end{abstract}


\maketitle

Quantum coherence is deemed responsible for a large variety of features, ranging from fundamental physical effects such as super-fluidity,
via a broad range of counter-intuitive interference and correlation phenomena with potential implications in the realm of quantum information technologies \cite{chuang00}
to transport processes  \cite{Lattices1956} even at the mesoscopic scale on the border between physics, chemistry and biology \cite{Engel2007}.
A central coherence property of {\it e.g.} a photonic wave packet is its coherence length,
but the extension of such a concept to composite quantum systems is by no means straight forward.

Entanglement theory promises an accurate characterization of coherence properties in multi-partite quantum systems in terms of $k$-partite entanglement, {\it i.e.} the minimum number $k$ of entangled components necessary to describe an $n$-partite system (in literature also referred to as depth of entanglement \cite{PhysRevLett.86.4431} or $k$-producibility \cite{Guhne2005}).
The definition of these concepts (as given in Eq.~\eqref{eq:Kent} below) is rather elementary, but, due to its non-constructive nature, the identification of $k$-partite entanglement in given mixed quantum states is a largely open problem:
up to now the theory of bipartite entanglement ({\it i.e.} $k=2$) has been developed fairly well \cite{PhysRevLett.80.2245,*PhysRevLett.77.1413,*Horodecki1996,*PhysRevA.65.032314},
and there has been substantial progress in the identification of genuine $n$-partite entanglement ({\it i.e.} $k=n$) \cite{Horodecki2001,*Guhne2009,Guhne2010,PhysRevLett.104.210501}. On the scales in-between, for $n>k>2$, however, only punctual knowledge, typically for states of specific type or size, is currently available \cite{PhysRevLett.86.4431,Guhne2005,Papp2009}.

The ability to probe these scales in-between is highly desirable for various reasons:
while it is well established that quantum computations with pure states necessarily require a large amount of entanglement in order to perform beyond the classically achievable \cite{Jozsa2003},
the situation is not as evident for mixed states as they would occur in realistic implementations, since also mixed separable states can lead to improved computational power \cite{Braunstein1999,Datta2008}. The possibility to identify entanglement properties in a more fine-grained version than currently possible for the mixed case would certainly help to  understand which specific features of multi-partite quantum states are really necessary for the appraised quantum speed-up.

In precision interferometry, the full enhancement of precision based on $n$ particles can be obtained only for a genuinely $n$-partite entangled state \cite{Giovannetti2011}. Entanglement between fewer components will result in a precision closer to the achievable with $n$ independent particles: identifying the largest $k$-partite entanglement (for $k$$\leq$$n$) that can be realised at given experimental conditions provides therefore very rigorous limitations to the achievable precision.

Similarly, such an assessment permits to estimate the number of nodes over which coherence in a computational network has been achieved \cite{Choi2010}. Fast excitation transport through molecular- or spin- networks has been shown to be associated with quantum coherence between an intermediate number of nodes \cite{PhysRevE.83.021912}, and such coherence can be identified through $k$-partite entanglement after projection onto the single-excitation subspace \cite{Tiersch2012}. This provides a very accurate characterisation of the spatial extent over which a multi-partite system displays quantum mechanical features, and the environmentally induced degradation of coherence can then be followed to monitor the emergence of classicality in a rather detailed fashion.

Our goal in the present contribution is, therefore, to provide for any system size $n$ a full hierarchy of separability criteria to characterize  multi-partite entanglement:
the criterion at the top of each hierarchy identifies genuine $n$-partite entanglement, followed by criteria that are positive only for states with at least $k$-partite entanglement for $k$ ranging from $n-1$ to $2$.

\begin{figure}
\includegraphics[width=0.4\textwidth]{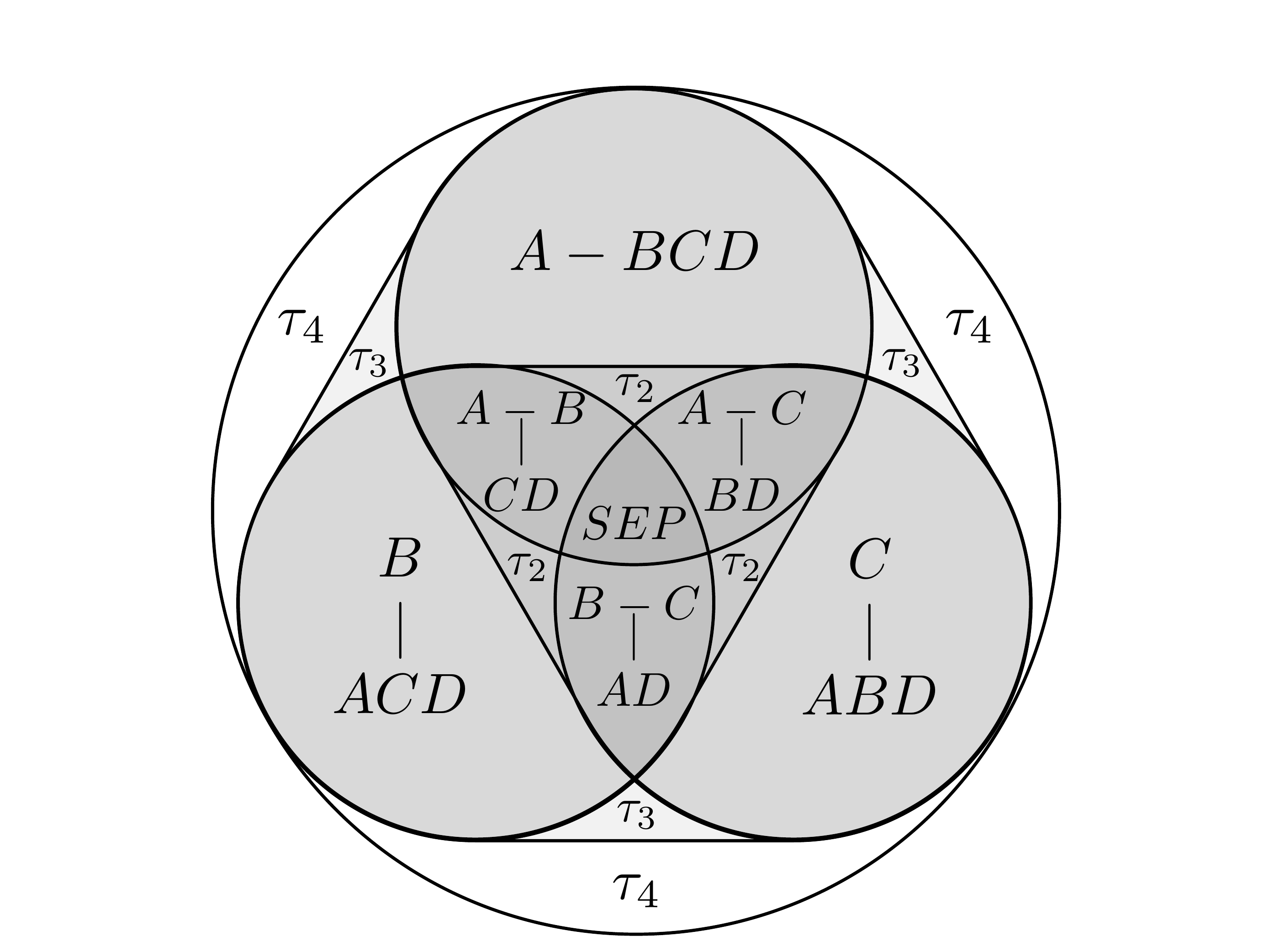}
\caption{\label{fig:eggs}
Simplified, schematic structure of $4$-partite states:
the grey circles depict quantum states that are separable with respect to three different bipartitions  (the biseparation $ABC-D$ and biseparations into pairs of subsystems are not shown.).
States that belong to two of these sets are at most bipartite entangled, and states that belong to all three sets are separable.
Any convex sum of bipartite entangled states (depicted by $\tau_2$) is considered at most bipartite entangled, even though the state might not be separable with respect to any bipartition. Similarly, any convex sum of tripartite entangled states (depicted by $\tau_3$) is considered at most tripartite entangled.
Only states that can not be obtained as a convex sum of at most tripartite entangled states are four-partite entangled.}
\end{figure}

Before introducing our framework, let us review briefly the necessary formal background.
A pure state $\ket{\Psi_{n,n}}$ of an $n$-partite quantum system is considered $n$-partite entangled if there is no separation of the sub-systems into two groups, such that $\ket{\Psi_{n,n}}$ could be described as the tensor product of  states of these two groups.
Analogously, an $n$-partite state $\ket{\Psi_{k,n}}$ is considered $k$-partite entangled if it can not be described without an at least $k$-partite entangled contribution.
If a state is not at least bipartite entangled, then it is separable.
For pure states, definition and identification of $k$-partite entanglement is rather straight forward,
but the situation changes drastically for mixed states:
a mixed $n$-partite state is considered $k$-partite entangled if it can not be expressed as a statistical mixture
\be
\sum_i\sum_{j=1}^{k-1}p_{ij}\prj{\Psi_{j,n}^{(i)}}\neq\varrho_{k,n}\ ,
\label{eq:Kent}
\ee
of at most ($k-1$)-partite entangled states with $p_{ij}\ge 0$ \cite{Guhne2005}.
This leads to a rather intricate structure of multi-partite entanglement as sketched in Fig.~\ref{fig:eggs}.

The task of our present hierarchies of separability criteria is to provide a potentially accurate identification of $k$-partite entanglement in mixed states. We first start out describing the underlying idea in rather general terms, followed by a specific realization that satisfies all of the desired properties.
What we aim at is a set of functions
\be
\tau_{k,n}(\varrho)=f(\varrho)-\sum_{i=1}^{n/2}a_{i}^{(k,n)}\sum_jf_{ij}(\varrho)
\label{eq:taugeneral}
\ee
defined in terms of functions $f$ and $f_{ij}$ where the index $j$ labels all inequivalent bipartitions \footnote{To any $i$-bipartition, there is an equivalent ($n$-$i$)-bipartition. In particular for $i=n/2$ some care is necessary to avoid double-counting.}
of the $n$-partite system in an $i$-partite and an ($n-i$)-partite component, referred to as $i$-bipartitions in the following.
Furthermore, the functions $f$ and $f_{ij}$, and the scalar weight factors $a_{i}^{(k,n)}$ need to satisfy the following conditions

\begin{itemize}
\item[I]$\tau_{k,n}(\varrho)$ is convex, {\it i.e.} $\sum_ip_i\tau(\varrho_i)\ge\tau(\sum_ip_i\varrho_i)$\label{ena}
\item[II]$f(\varrho)\ge 0$ and $f_{ij}(\varrho)\ge 0$, $\forall i,j,\varrho\ge 0$.
\item[III]$f(\psi)=f_{ij}(\psi)$ if $\ket{\psi}$ is bi-separable with respect to the $j$-th $i$-bipartition.
\item[IV]$a_{i}^{(k,n)}\ge 0$, $\forall i,k,n$.
\end{itemize} 

Convexity of $\tau_{k,n}$ allows us to restrict the following discussion to pure states: if $\tau_{k,n}$ is non-positive for all pure states with less than $k$-partite entanglement, condition I entails that a positive value of $\tau_{k,n}$ identifies (at least) $k$-partite entanglement in mixed states.
What remains to be done is to tailor the prefactors $a_{i}^{(k,n)}$ in order for $\tau_{k,n}$ to have the desired properties for pure states.
Since $f_{ij}(\Psi)$ coincides with $f(\Psi)$ if $\ket{\Psi}$ is separable with respect to the $j$-th $i$-bipartition, it is sufficient to characterize the separability properties of pure $k$-partite entangled $n$-partite states,
and choose the weights $a_i^{(k,n)}$ such that
$\sum_{ij|bs}a_{i}^{(k,n)}f_{ij}(\Psi_{k^\prime n})\ge f(\Psi_{k^\prime n})$ for any $k^\prime$-partite entangled state with $k^\prime<k$, where the sum runs over all bipartitions with respect to which $\ket{\Psi_{k^\prime n}}$ is separable.
Due to the positivity of $f_{ij}$ and $a_i^{(k,n)}$ this directly implies that $\tau_{k,n}$ is non-positive for all states which are not at least $k$-partite entangled

\begin{figure}
\includegraphics[width=0.4\textwidth]{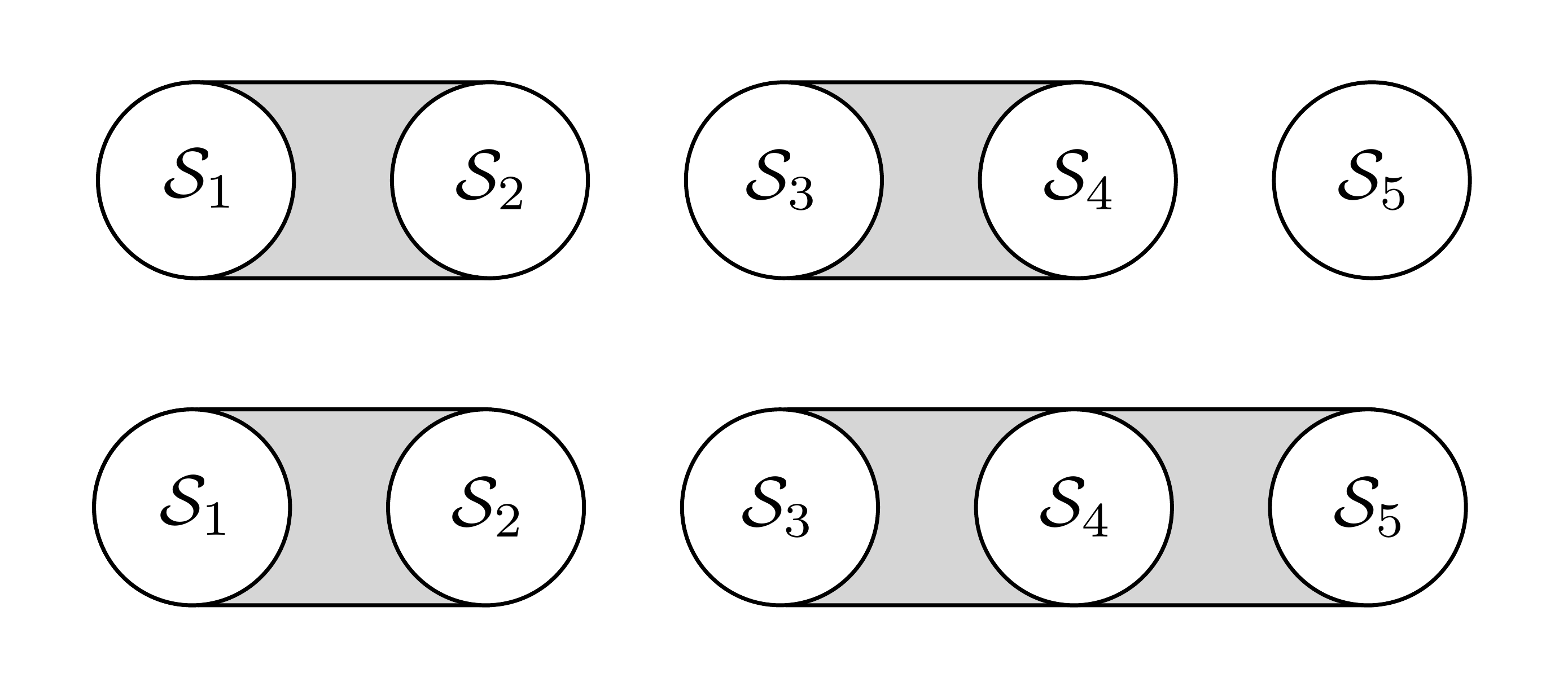}
\caption{\label{fig:knexp}
Schematic representation of two pure five-partite states: a bipartite entangled (top) and a tripartite entangled (bottom).
The bipartite entangled state is separable with respect to the 2-bipartitions ${\cal S}_1{\cal S}_2-{\cal S}_3{\cal S}_4{\cal S}_5$ 
and ${\cal S}_1{\cal S}_2{\cal S}_5-{\cal S}_3{\cal S}_4$.
Indeed, any pure bipartite entangled five-partite state is biseparable with respect to at least two 2-bipartitions.
Pure tripartite entangled five-partite states on the other hand can be separable with respect to one 2-bipartition only.}
\end{figure}

As indicated in Fig.~\ref{fig:knexp}, pure states with only a small entangled component are biseparable with respect to many bipartitions, so that many components $f_{ij}(\Psi)$ coincide with $f(\Psi)$ and the weights can be chosen comparatively small.
Choosing the weight factors $a_{i}^{(k,n)}$ increasing with $k$ will thus allow us to arrive at the desired hierarchies.

At the bottom of the hierarchies lies $\tau_{2,n}$ which has to be non-positive for all completely separable states $\ket{\Psi_{1,n}}$.
Since $\ket{\Psi_{1,n}}$ is separable with respect to any bipartition, we have $f(\Psi_{1,n})=f_{ij}(\Psi_{1,n})$ $\forall i,j$ due to (II).
Any choice satisfying $\sum_{i}a_{i}^{(2,n)}=1$ will therefore result in $\tau_{2,n}(\Psi_{1,n})=0$ for any completely separable state $\ket{\Psi_{1,n}}$.
In order to proceed we need to tailor the $a_{i}^{(3,n)}$ such that $\tau_{3,n}$ is non-positive for all pure states that contain less than tripartite entanglement.
As depicted in Fig.~\ref{fig:knexp} with the exemplary case of $n=5$, any pure state $\ket{\Psi_{2,n}}$ (for $n\neq 4$)
 is separable with respect to at least $m$ $2$-bipartitions, where $m$ is the largest integer $\le n/2$  \footnote{For $n=4$ these two 2-bipartition are equivalent,
 so that $a_{i}^{(3)}=\delta_{i2}$}.
Accordingly, $a_{i}^{(3,n)}=\delta_{i2}/m$ is a valid choice for $\tau_{3,n}$. 

Typically there is not a unique choice for the weights $a_{i}^{(k,n)}$, and the resulting freedom can be used to optimize the functions $\tau_{k,n}$ for specific quantum states.
As a rough rule of thumb we found that for states with highly mixed reduced density matrices choices with large weight factors $a_{i}^{(k,n)}$ for $i\simeq n/2$ and small or vanishing ones for $i\ll n/2$ yield strong criteria.
For example for odd $n>5$,
$a_{i}^{(3,n)}=\delta_{i3}/m$ and
$a_{i}^{(3,n)}=\delta_{i2}/m$
are both valid choices to define $\tau_{3,n}$, but we found the former to result in a stronger criterion.
Similarly, for $n>8$ $a_{i}^{(3,n)}=\delta_{i4}/(m(m-1)/2)$ typically leads to an even stronger criterion.
Since picking a good choice for the weight factors helps to identify good criteria,
we refrain from providing a systematic description for the construction of the $a_i^{(k,n)}$,
but rather depict choices for $n\le 7$ that we found to yield good results in Table~\ref{table}.
The functions $\tau_{k,n}$ with these specific coefficients then define a full hierarchy of necessary separability criteria for any system size $n$.

\begin{table*}
\rowcolors{1}{}{}
\begin{tabular}{|c||c|c|c|c|c|c|c||c|}
\hline
&$N=12$&$N=11$&$N=10$&$N=9$ & $N=8$&$N=7$&$N=6$&\\
\hline\hline
$K=2$&[0,0,0,0,0,$1/462$]&[0,0,0,0,$1/462$]&$[0,0,0,0,1/126]$&$[0,0,0,1/126]$&[0,0,0,$1/35]$&[0,0,$1/35$]&[0,0,$1/10$]&\\
\hline
$K=3$&[0,0,0,0,0,1/10]&[0,0,0,0,1/10]&$[0,0,0,1/10,0]$&$[0,0,0,1/6]$&$[0,0,0,1/3]$&[0,0,1/3]&[0,1/3,0]&\\
\hline
$K=4$&[0,0,0,0,0,1/3]&[0,0,0,0,1/3]&$[0,0,0,1/3,0]$&$[0,0,1/3,0]$&$[0,0,1/2,1/3]$&$[0,0,1/2]$&[0,1/3,1]&\\
\hline
$K=5$&[0,0,0,1/3,0,1/3]&[0,0,0,1/2,1/3]&$[0,0,0,1/2,0]$&$[0,0,1/3,1/2]$&$[0,0,1/2,1]$&[0,0,1]&[0,1,1]&\\
\hline
$K=6$&[0,0,0,1/3,1/2,1/3]&[0,0,0,1/2,1/2]&$[0,0,0,1/2,1]$&$[0,0,1/3,1]$&[0,0,1,1]&[0,1,1]&[1,1,1]&\\
\hline 
$K=7$&[0,0,0,1/3,1/2,1]&[0,0,0,1/2,1]&[0,0,0,1,1]&[0,0,1,1]&[0,1,1,1]&[1,1,1]&\cellcolor[gray]{0.9}&\\
\hline
$K=8$&[0,0,0,1/3,1,1]&[0,0,0,1,1]&[0,0,1,1,1]&[0,1,1,1]&[1,1,1,1]&\cellcolor[gray]{0.9}&[1,1]&$K=5$\\
\hline
$K=9$&[0,0,0,1,1,1]&[0,0,1,1,1]&[0,1,1,1,1]&[1,1,1,1]&\cellcolor[gray]{0.9}&[1,1]&[0,1]&$K=4$\\
\hline
$K=10$&[0,0,1,1,1,1]&[0,1,1,1,1]&[1,1,1,1,1]&\cellcolor[gray]{0.9}&[1]&[0,1]&[0,1/2]&$K=3$\\
\hline
$K=11$&[0,1,1,1,1,1]&[1,1,1,1,1]&\cellcolor[gray]{0.9}&[1]&[1/3]&[0,1/3]&[0,1/10]&$K=2$\\
\hline
$K=12$&[1,1,1,1,1,1]&\cellcolor[gray]{0.9}&&$N=2$&$N=3$&$N=4$&$N=5$&\\
\hline
\end{tabular}
\caption{Specific choices for the weight factors $a_i^{(KN)}$ that define valid criteria $\tau_{KN}$ to detect $K$-body entanglement in an $N$-body system.
Each vector in the table contains the elements $[a_{N/2}^{(KN)},\hdots, a_1^{(KN)}]$ (for even $N$) and  $[a_{(N-1)/2}^{(KN)},\hdots, a_1^{(KN)}]$ (for odd $N$).
The upper left half ranges from $N=12$ to $N=6$, the lower right from $N=2$ to $N=5$.
\label{table}}
\end{table*}


As is the case for any attempt to detect entanglement beyond $2\times 3$-dimensional systems \cite{Horodecki1996}, a tool can either identify entanglement or it can identify separability, but there is none that can assert with certainty whether a state is entangled or separable.
Also here a non-positive value of $\tau_{k,n}$ does not necessarily imply that the considered state was not $k$-partite entangled,
but it could also be due to the fact that $\tau_{k,n}$ is not strong enough to identify the targeted entanglement in the specific state.
If the latter is the case, one can improve $\tau_{k,n}$ provided there are additional properties that can be exploited.
Here we would like to demonstrate this with the example of $W$-states \cite{PhysRevA.62.062314}, {\it i.e.} states with a single excitation $\ket{W}=\sum_iw_i\ket{i}$, where $\ket{i}$ is a short hand notation for the state with the $i$-th subsystem in its excited state and all other subsystems in their ground state.
These states attract particular attention since they occur naturally in excitation transport processes \cite{Plenio2008,*Rebentrost2009b,PhysRevE.83.021912},
and they also permitted the observation of genuine multi-partite entanglement of an eight ion string \cite{Haffner2005}.

If such a $W$-state is bi-separable with respect to an $i$-bipartition, so that $\ket{W}=\ket{\phi_1}\otimes\ket{\phi_2}$,
then one of the components $\ket{\phi_{1/2}}$ needs to be completely separable, because otherwise there would be a finite amplitude for two excitations \footnote{In terms of the expansion coefficients $w_i$, factorization of $\ket{W}$ into a $i$-partite component $\ket{\phi_1}$ and an $n-i$-partite component $\ket{\phi_2}$ implies that either $i$ or $n-i$ coefficients vanish.}. Consequently, biseparability with respect to an $i$-bipartition ($i\le n/2$) implies biseparability with respect to at least $i$ one-bipartitions, $i(i-1)/2$ two-bipartitions, and similarly for larger bipartitions.
These additional separability properties permits to identify significantly lower values for the weights $a_{i}^{(k,n)}$ than those for general states as given in Table~\ref{table}.
In contrast to above, where we found strong criteria based on bipartions of $\simeq n/2$ subsystems, in the case of $W$-states it is rather advantageous to focus on 1-bipartitions: the weights $a_i^{(k,n)}=\delta_{i1}/(n-(k-1))$ for $k\neq2$ and $a_i^{(k,n)}=\delta_{i1}/n$ for $k=2$ provide a strong hierarchy $\tau_{k,n}^w$ for $W$-states.
In a similar fashion, the present criteria can also be adjusted for different classes of states, such as more general Dicke states \cite{PhysRev.93.99} or potentially states with permutation symmetries \cite{Toth2009}.

So far we have discussed the hierarchies in a rather abstract setting, assuming the existence of functions that satisfy the above list of properties I to IV.
Let us become more specific now and present a possible choice of such functions.
It is based on the fact that a twofold tensor-product $\ket{\Psi}\otimes\ket{\Psi}$ of a state with itself features very specific invariance properties if $\ket{\Psi}$ is not genuinely $n$-partite entangled \cite{MINTERT2005}:
$\ket{\Psi}$ is biseparable with respect to a biseparation that divides the system in the components $A$ and $B$ if and only if
the two-fold state $\ket{\Psi}\otimes\ket{\Psi}$ is invariant under the permutation that permutes the two $A$-components (or, analogously, the $B$-components).
Taking $f$ to be a function $g$ of $\ket{\Psi}\otimes\ket{\Psi}$ and $f_{ij}=g(\Pi_{ij}\ket{\Psi}\otimes\ket{\Psi})$ where $\Pi_{ij}$ is the permutation that permutes the $A$-components associated with the $j$-th $i$-bipartition makes sure that condition III is satisfied.
Condition I, {\it i.e.} convexity of $\tau_{k,n}$, is in general difficult to achieve,
but
\bqa
f&=&\sqrt{\matel{\Phi_S}{{\bf \Pi}\varrho\otimes\varrho}{\Phi_S}}\ \mbox{and}\nonumber\\
f_{ij}&=&\sqrt{\matel{\Phi_S}{\Pi_{ij}\varrho\otimes\varrho\Pi_{ij}^\dagger}{\Phi_S}}
\label{eq:tau}
\eqa
with the global permutation ${\bf \Pi}$ and a product vector $\ket{\Phi_S}$, are convex resp. concave \cite{PhysRevLett.104.210501,Guhne2010}.

For pure states $f$ coincides with $\sqrt{\matel{\Phi_S}{\varrho\otimes\varrho}{\Phi_S}}$ ($\ket{\Psi}\otimes\ket{\Psi}$ is invariant under ${\bf\Pi}$), so that condition III is satisfied,
and the present specific choices for $f$ and $f_{ij}$ are indeed non-negative.
As long as $a_{i}^{(k,n)}$ are non-negative as they should be according to condition IV,
Eqs.~\eqref{eq:taugeneral} and \eqref{eq:tau} with the weight factors $a_i^{(k,n)}$ such as those given in Table~\ref{table} provide a valid realization of a hierarchy following conditions I through IV.

We have tested the performance of these hierarchies for different, exemplary cases, comparing it with previously known criteria \cite{PhysRevLett.86.4431,Lougovski2009}  for 4- and 6-partite spin-squeezed states and 4-partite W-states. This comparison is shown in the section A of the appendix. This test demonstrates how the hierarchies $\tau_{k,n}$ and $\tau_{k,n}^w$ often outperform prior techniques, especially in presence of strong mixing.
This is remarkable, in particular, since existing criteria have been specifically tailored to address entanglement properties of a given class of states,
whereas we just varied the coefficients $a_i^{(k,n)}$ retaining the same analytic form.
By a systematic application of the hierarchy $\tau_{k,n}$ to a given system of interest it is possible to gain a deep insight in its entanglement structure,
as noticeable for the exemplary case of spin-squeezed states in the appendix, where a rich underlying multi-partite entanglement landscape is uncovered in a broad interval of the spin-squeezing parameter where previously only bipartite entanglement was detected.

As argued above, the possibility to detect $k$-partite entanglement for varying $k$ also provides a refined insight in entanglement dynamics. This is substantiated in Sec B of the appendix with the investigation of a fully connected 12-partite graph state undergoing a dephasing evolution, where it is emphasized how the different types of $k$-partite entanglement decay on different time scales.

The possibility to explore the dynamics of arbitrary $k$-partite entanglement in turn enables to uncover relevant physical features of a given system,
as we exemplify with the verification of three-body interactions \cite{Mizel2004} in section C of the appendix.

These examples underline the usefulness of the present hierarchies; the specific framework of permutation operators that we have used for the explicit construction of the present hierarchy is by no means the sole way to arrive at such a hierarchy. Many other typically employed tools, such as entanglement witnesses \cite{PhysRevLett.106.190502} or positive maps \cite{Peres1996} bear potential for a systematic construction.
Also, whereas we have focussed here on the classification of $k$-partite entanglement, a classification in more refined classes according to LOCC-inequivalence \cite{PhysRevA.62.062314,PhysRevA.65.052112} seems feasible.

Inspiring discussions with \L ukasz Rudnicki, Otfried G\"uhne and generous financial support by the European Research Council are gratefully acknowledged. 

\section{Appendix}

This appendix shows various applications of the hierarchies $\tau_{k,n}$ in order to provide a quantitative estimation of their performance and to exemplify their potential uses. In the first section we compare the strength of the hierarchies with previously introduced separability criteria \cite{Lougovski2009,PhysRevLett.86.4431} for W-states in \ref{sec:wstates} and spin-squeezed states in \ref{sec:spinsqueeze}. Subsequently, in section \ref{sec:deco}, we utilise one hierarchy to investigate the decay of multi-partite entanglement for the case of a genuine $12$-partite entangled state undergoing a dephasing dynamics. Furthermore in section \ref{sec:slopes} we demonstrate how detecting $k$-partite entanglement in an $n$-partite system with $k\neq n$ provides a tool to distinguish between three and two body interactions.

\subsection{Comparison with previous criteria}
\subsubsection{W-states} \label{sec:wstates}

A criterion to detect $k$-partite entanglement in $4$-partite $W$-states, {\it i.e.} states with a single excitation, has been described in \cite{Lougovski2009} based on expectation values of a given density matrix $\varrho$ with respect to the four states
\begin{eqnarray}
\begin{aligned}
|W_1\rangle=\frac{1}{2}(|1000\rangle+\mathrm{e}^{i\phi_1}|0100\rangle+\mathrm{e}^{i\phi_2}|0010\rangle+\mathrm{e}^{i\phi_3}|0001\rangle)\ ,\\
|W_2\rangle=\frac{1}{2}(|1000\rangle-\mathrm{e}^{i\phi_1}|0100\rangle-\mathrm{e}^{i\phi_2}|0010\rangle+\mathrm{e}^{i\phi_3}|0001\rangle)\ ,\\
|W_3\rangle=\frac{1}{2}(|1000\rangle+\mathrm{e}^{i\phi_1}|0100\rangle-\mathrm{e}^{i\phi_2}|0010\rangle-\mathrm{e}^{i\phi_3}|0001\rangle)\ ,\\
|W_4\rangle=\frac{1}{2}(|1000\rangle-\mathrm{e}^{i\phi_1}|0100\rangle+\mathrm{e}^{i\phi_2}|0010\rangle-\mathrm{e}^{i\phi_3}|0001\rangle)\ .
\end{aligned}
\end{eqnarray}
As shown in \cite{Lougovski2009}, the variance based quantity
\begin{equation}
\label{eq:delta}
\Delta(\varrho)=1-\sum_{i=1}^4\langle W_i|\varrho|W_i\rangle^2,
\end{equation}
permits to identify bi-, tri- and four-partite entanglement:
if $\Delta(\varrho)<\Delta_k$ for some choice of the phases $\phi_1$, $\phi_2$ and $\phi_3$, with $\Delta_2$$=$$3/4$, $\Delta_3$$=$$1/2$ and $\Delta_4$$=$$5/12$,
then $\varrho$ is $k$-partite entangled.

In order to compare the entanglement criteria we generated ensembles of random W-states $\varrho(A)=AA^\dagger/\mbox{Tr}(AA^\dagger)$ in terms of random matrices $A$ whose elements are independently and normally distributed. This ensures a distribution according to the Hilbert-Schmidt measure \cite{Zyczkowski2001}, and the distribution of the entropies of the states $\rho(A)$ is determined by the ratio of the mean $\bar \mu$ and variance $\sigma_\mu$ of the Gaussian distribution from which the matrix elements are obtained.
With states generated in this fashion we will compare in the following
\be
d_k(\varrho)=\min_{\phi_1,\phi_2,\phi_3}\Delta(\varrho)-\Delta_k ,
\ee
{\it i.e.} the criterion from \cite{Lougovski2009} with an optimal choice of phases $\phi_i$, with the hierarchy $\tau^w_{k,n}$, where we also optimise over the separable states $|\Phi_S\rangle$ introduced in Eq.(ef{eq:tau}).

\begin{figure}
\includegraphics[width=0.4\textwidth]{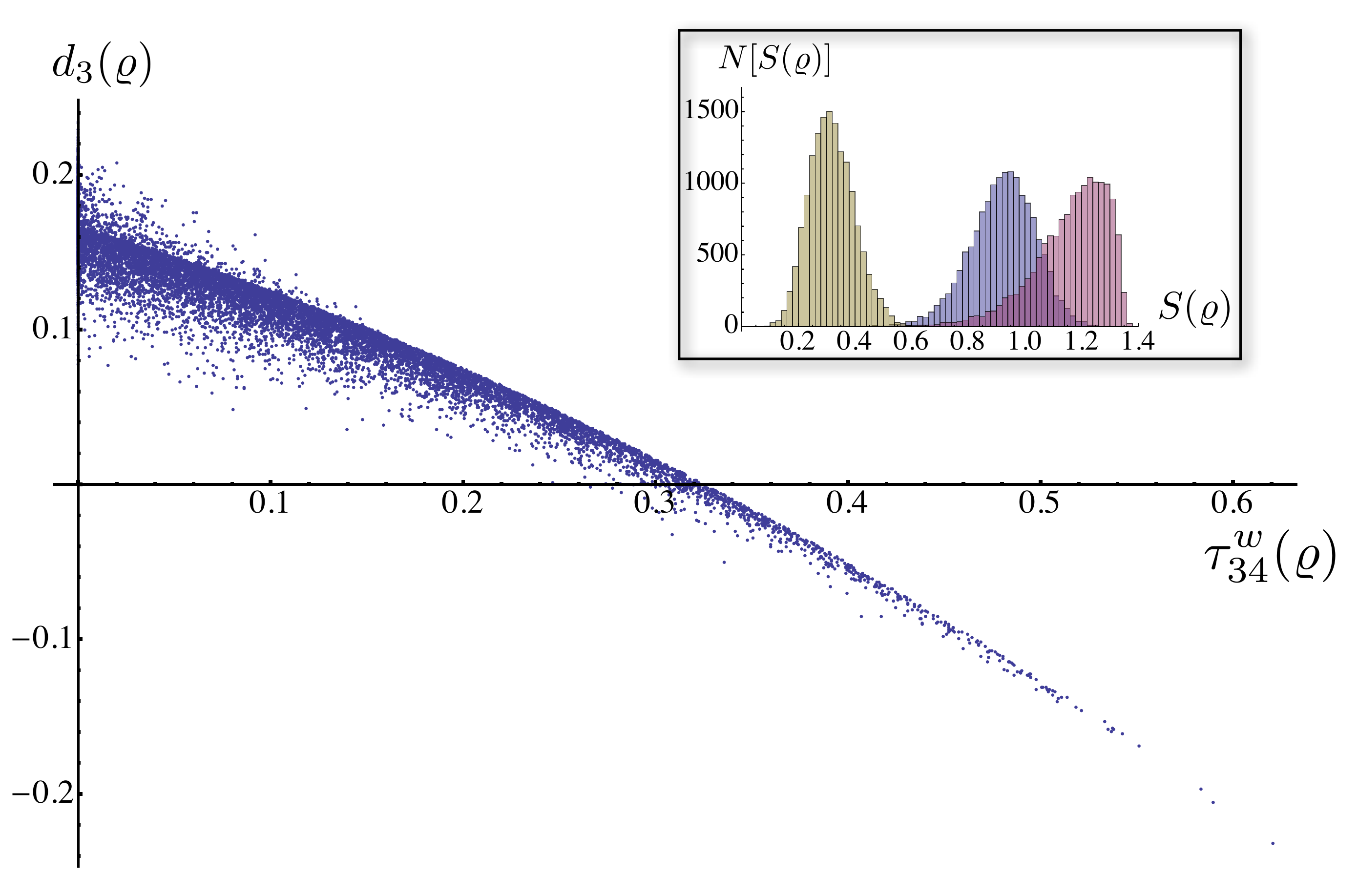}
\caption{\label{fig:wcompare}
Comparison between the hierarchy $\tau^w_{k,4}$ and bound Eq.\eqref{eq:delta} for the case $k$=3. In the inset the distribution of entropies for the three different batches of data are depicted. The main picture shows obtained values for the two criteria with states whose entropy follows the blue (middle) distribution in the inset, each point representing a single random mixed state. All states under the $y=0$ axis violate bound Eq.(\ref{eq:delta}) and are therefore detected by $d_3$. Tripartite entanglement is detected by the presented hierarchy when $\tau^w_{3,4}$ is positive. No state with $d_3$$<$0 leads to a non-positive value of $\tau^w_{3,4}$. The majority ($\sim$60\%) of random states, especially for higher mixedness, yield a positive $\tau^w_{3,4}$, but a variance $\Delta(\rho)$ not small enough to detect tripartite entanglement according to Eq.(\ref{eq:delta}). The value for $\tau^w_{3,4}$ has been normalised to unity for a perfect W state.}
\end{figure}

The choices $\bar \mu=0$ and $\sigma_\mu=1$ lead to an ensemble of intermediate mixing with an entropy distribution shown in blue in the inset of Fig.~\ref{fig:wcompare}.
All states are detected as bipartite entangled according to both $d_2$ and $\tau^w_{2,4}$, regardless of their mixedness.
For the case of tripartite entanglement, however, there is a rather striking difference in the performance of the two criteria as it can be seen in Fig.~\ref{fig:wcompare}, where $d_3$ is plotted against $\tau^w_{3,4}$ for this ensemble.
Whereas $d_3$ detects only $5\%$ of the ensemble members as tripartite entangled, $\tau^w_{3,4}$ manages to detect $65\%$,
and there is no state detected by $d_3$ that is not detected by $\tau^w_{3,4}$.

In the case of genuine $4$-partite entanglement it had already been observed that $\tau^w_{4,4}$, \textit{i.e.} the genuine multipartite entanglement criteria presented in \cite{Huber2010}, is not particularly strong for $W$-states, and, indeed, there are various cases where $d_4$ performs better than $\tau^w_{4.4}$: 0.05\% states yield a negative value of $d_4$ whereas only 0.003\% lead to positive $\tau^w_{4,4}$. This can however be accounted for by a modification of $\tau^w_{4,4}$ (Eq.(III) in \cite{Huber2010}) that is better suited for $W$-states. With this modification the topmost members of both hierarchies detect 0.05\% of all states and agree in more than 90\% of the detected cases.

Inspecting the entropies of the states that are detected by $\tau^w_{3,4}$ but not by $d_3$, one realizes that $\tau^w_{3,4}$ performs better in particular for rather highly mixed states.
We therefore considered two additional ensembles: with $\bar \mu=3\, (20)$ and $\sigma_\mu=2\,(8)$ what results in rather high (low) entropies as depicted in red (yellow) \footnote{In order to increase even further the mixedness of the distribution generated with $\bar \mu=3$ and $\sigma_\mu=2$, the diagonal elements of the density matrices have been multiplied by 10. This ensemble is no longer strictly randomly distributed according to the Hilbert-Schmidt measure as in the other two cases, but permits a good inspection of the behaviour of very highly mixed states}  in the inset in Fig.~\ref{fig:wcompare}. For close to pure states (yellow distribution) both criteria perform equally well, but in the case of highly mixed states (red distribution), the situation changes drastically: $d_3$ does not detect any state, whereas $\tau^w_{3,4}$ still manages to identify $5\%$ and thus proves to be significantly stronger for highly mixed states.

To better explore the different performances of the two criteria depending on the mixing of the quantum states we consider now a Werner-type state
\begin{equation}
\varrho_q=q|W_1\rangle\langle W_1|+(1-q)\frac{\mathbb{1}_n}{2^n},
\end{equation}
in order to systematically explore mixedness via a single parameter $q$. With both inequality Eq.\eqref{eq:delta} and the hierarchy $\tau^w_{k,n}$, $4$-partite entanglement is detected for $q$$>$2/3, and bipartite entanglement for all values of $q$$>$0. However, the threshold value for tripartite entanglement detection is found to be $q\approx$ 0.58 by violation of Eq.\eqref{eq:delta}, and $q$=0.3 through the hierarchy. That is, $\tau^w_{3,4}$ detects nearly twice the parameter range to be tripartite entangled than $d_3$.

\subsubsection{Spin-squeezed states} \label{sec:spinsqueeze}

Also for spin-squeezed states \cite{Kitagawa1993} a criterion for the identification of $k$-partite entanglement has been described before. Spin-squeezed states are entangled, and the strength of spin-squeezing is related the degree of entanglement \cite{PhysRevLett.86.4431}. More specifically, the collective spin of a composite system comprised of $n$ spin-$1/2$ particles which is $k$-partite entangled ($k$=1 means separable) satisfies the following inequality
\begin{equation}
\label{eq:boundspin}
\mathrm{Var}(\mathcal{J}_z)\geq \frac{n}{2}F_{k/2}\left(\frac{\langle\mathcal{J}_x\rangle}{n/2}\right)\ ,
\end{equation}
where $k$ is a divisor of $n$,
$\mathcal{J}_x$ and $\mathcal{J}_z$ are collective spin operators and $F$ is a set of numerical bounds that can be found in \cite{PhysRevLett.86.4431}. Violation of Eq.\eqref{eq:boundspin} constitutes a sufficient criterion for $k+1$-partite entanglement, and we compare its performance with the hierarchy of criteria $\tau_{k,n}$ given by the choice of coefficients $a_i^{(k,n)}$ listed in Table~\ref{table}.

\begin{figure*}
\includegraphics[width=0.8\textwidth]{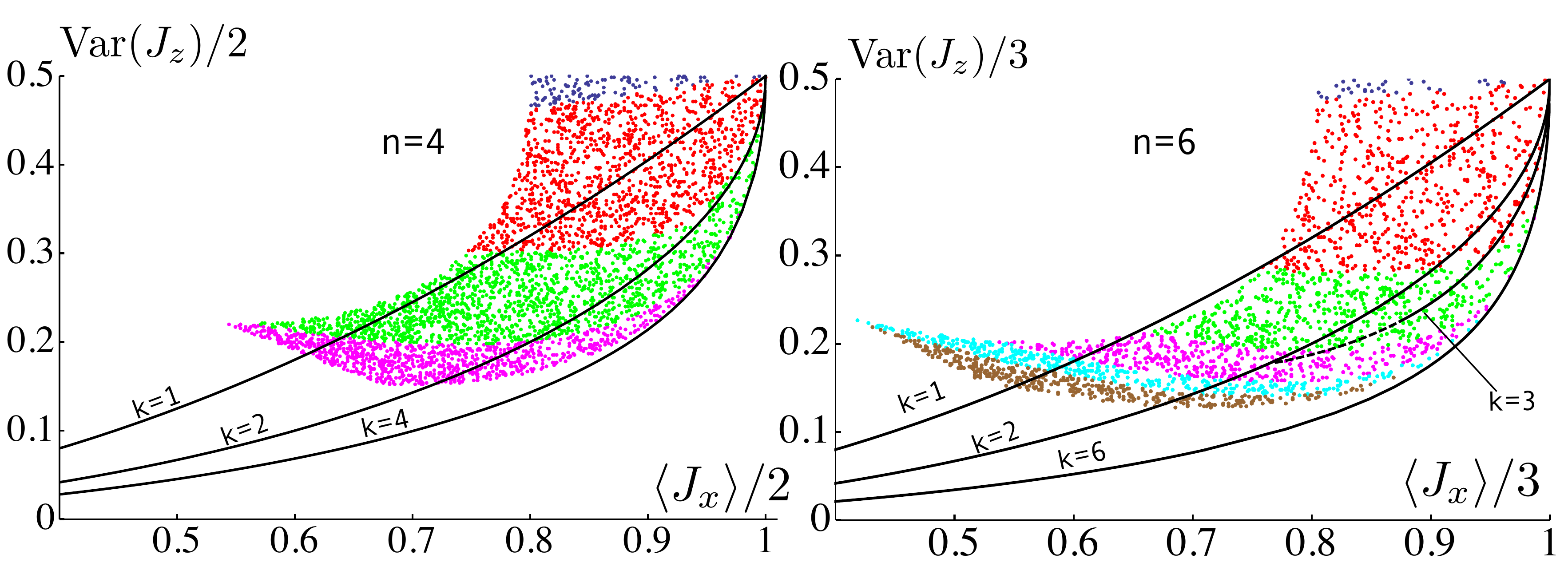}
\caption{\label{fig:spincompare2}
Comparison of the spin-squeezing inequality Eq.(\ref{eq:boundspin}) and the hierarchy $\tau_{k,n}$ for the case of $n$=4 (left) and $n$=6 (right) spin 1/2 particles. Each point corresponds to a single random state. The solid curves represent the bound Eq.(\ref{eq:boundspin}) for $k$=1,2 and 4 ($n$=4) and 1,2,3 and 6 ($n$=6) respectively from top to bottom. Lying below a certain $k$ curve gives a sufficient condition for $k+1$-partite entanglement. Colour-code: states which lead to a $\tau_{k,n}>0$ are brown ($k$=6), cyan ($k$=5), magenta ($k$=4), green ($k$=3), red ($k$=2); blue points correspond to states for which no entanglement has been detected. The curve for $k$=3 in $n$=6 cannot be rigorously generated for all values of $\langle J_x\rangle$: the solid line depicts where Eq.(\ref{eq:boundspin}) is saturated, and the dashed line its analytical continuation which overestimates the bound.}
\end{figure*}

Spin squeezed states are of the form
\be
\ket{\chi}=e^{-i\chi \mathcal{J}_z^2}\ket{\Phi_0}\ ,
\ee
where
\be
\ket{\Phi_0}=\left[\frac{1}{\sqrt{2}}(|0_z\rangle+|1_z\rangle)\right]^{\otimes n},
\ee
is the eigenstate of $\mathcal{J}_x$ to the eigenvalue $n$/2.
Since we are interested in mixed states, we consider the admixture of white noise, \ie Werner-type states:
\begin{equation}
\varrho_{\chi,\eta}=\eta\prj{\chi}+(1-\eta)\frac{\mathbb{1}_n}{2^n}\ ,
\end{equation}
with the $n$-partite identity $\mathbb{1}_n$.
For the present comparison we consider the cases $n=4(6)$ and generated 4000 (2500) random states each with $\chi$ and $\eta$ distributed uniformly in the intervals $0$$<$$\chi$$<$$0.5$ and $0.8$$<$$\eta$$<$$1$.
To achieve optimal performance of the spin squeezing inequality Eq.~\eqref{eq:boundspin},
expectation values of $\mathcal{J}_z$ are not taken with respect to $\varrho_{\chi,\eta}$, but with respect to $\mathrm{exp}(-i\nu \mathcal{J}_x)\,\varrho_{\chi,\eta}\,\mathrm{exp}(i\nu \mathcal{J}_x)$, and an optimisation over the angle $\nu$ is performed.

Fig.~\ref{fig:spincompare2} depicts the minimal variance Var$(\mathcal{J}_z$) as function of $\mathcal{J}_z$ for various values of $k$ with $n=4$ and $n=6$.
The black lines delimit the areas in which Eq.~\eqref{eq:boundspin} is violated, and $k+1$-partite entanglement is identified;
in some cases different degrees are not resolved due to the limitation that $k$ be a divisor of $n$.
Each of the colored points corresponds to one random state, and the color coding refers to the degree of entanglement that is detected by the positivity of $\tau_{k,n}$  (details in the caption of Fig.~\ref{fig:spincompare2}).

For $n=4$, the hierarchy $\tau_{k,n}$ is strictly better than Eq.~\eqref{eq:boundspin}: not only is it able to differentiate between tri- and four-partite entanglement, but within the set of states detected as bi(tri)partite entangled by the spin-squeezing inequality it shows that more than half of them ($\sim60\%\,(52\%)$) are actually $k$-partite entangled with $k>2(3)$. Furthermore, there are no states for which $\tau_{k,4}$ detects a smaller number of entangled components than as estimated via the spin-squeezing inequality.

Since for $n=6$ there are a few states where the violation of Eq.~\eqref{eq:boundspin} leads to an estimated $k$-partite entanglement larger than detected by the present hierarchy,
we shall analyze this case in more detail.
Out of all the states detected as bipartite entangled by $\tau_{2,6}$, and depicted in red in Fig.~\ref{fig:spincompare2},
$40\%$ are non-squeezed (they lie above the $k$=1 curve) and, therefore, can not be detected as entangled by Eq.~\eqref{eq:boundspin} as a matter of principle.
$53\%$ of the states depicted in red are detected as bipartite entangled also via their spin squeezing properties,
but for 7\% Eq.~\eqref{eq:boundspin}  detects a larger degree of entanglement than $\tau_{k,n}$ as depicted by the red dots below the $k$=2,3 curves.
Out of the states detected as tripartite entangled by the present hierarchy (depicted in green in Fig.~\ref{fig:spincompare2}),
$66\%$ lie between the $k$=1 and the $k$=2 curve and are detected only as bipartite entangled by Eq.~\eqref{eq:boundspin};
$\sim18\%$, however, are detected as four-partite entangled, since they lie below the $k$=3 curve.

About $52\% (49\%$ considering  only squeezed states) of the entire sample is detected as 4-partite, 5-partite or 6-partite entangled by $\tau_{k,n}$ (depicted by magenta, cyan and brown respectively in Fig.~\ref{fig:spincompare2}),
whereas Eq.~\eqref{eq:boundspin} does not permit to distinguish between 4-partite entanglement and $k$-partite entanglement with $k>4$.
In addition, a substantial portion of those states is detected by Eq.~\eqref{eq:boundspin} as bipartite entangled only, or is not detected as entangled at all. 

After all, Eq.~\eqref{eq:boundspin} and the present hierarchies result in inequivalent assessments, and the result of a comparison likely depends on the sample of states chosen.
Here we have been choosing the parameters such that there is a substantial portion of states with significant squeezing,
and found that Eq.~\eqref{eq:boundspin} identified a larger degree of entanglement in $\sim 7\%$ of the {\em squeezed} states,
whereas among $\sim 47\%$ of those states $\tau_{k,6}$ detects a larger number $k$ of entangled components.
That is, despite the fact that Eq.~\eqref{eq:boundspin} has been designed specifically for spin-squeezed states, and un-squeezed states have not been taken into account in the comparison,
our present hierarchies outperform spin-squeezing inequalities significantly more often than vice versa.

\subsection{Decoherence of open quantum systems}\label{sec:deco}

So far in this appendix the focus has been on comparing the detection strength of hierarchy $\tau_{k,n}$ with previously known criteria for specific classes of states. Now we aim to show the potential of the hierarchies to identify physical properties of multi-partite entanglement. This is exemplified by monitoring the qualitative changes of entanglement induced by decoherence in a multi-partite system undergoing a dephasing evolution.

\begin{figure}
\includegraphics[width=0.4\textwidth]{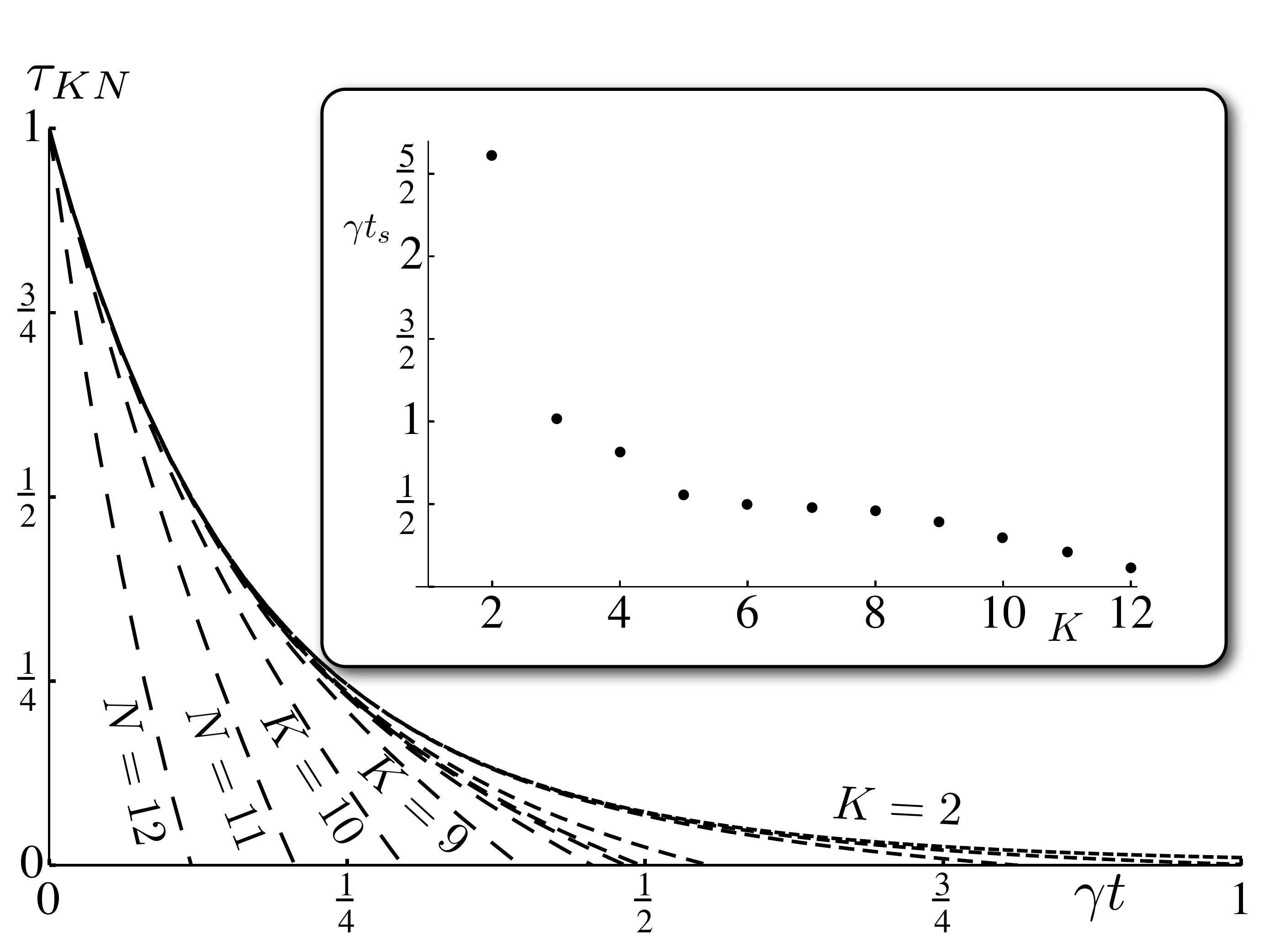}
\caption{\label{fig:decay}
Entanglement decay for an $n=12$-partite system initially prepared in a fully-connected graph state and evolving according to a dephasing channel for $k$ ranging from $2$ to $12$ depicted with increasingly long dashing.
The inset shows the instants $t_s$ from which no $k$-partite entanglement is detected anymore.
}
\end{figure}

To this end we choose a system of $n=12$ two-level systems and consider the impact of dephasing on a fully-connected graph state \cite{Hein2004}, where each subsystem is subject to a dephasing channel with Kraus operators
$E_1=\sqrt{(1+\exp (-\gamma t))/2}\ {\um}$ and $E_2=\sqrt{(1-\exp(-\gamma t))/2}\ \sigma_z$,
with $\gamma$ the single particle decoherence rate.
With these single-particle channels, the multi-partite dynamics reads 
\begin{equation}
\varrho(t)=\sum_{i_1\hdots i_{12}}E_{i_1}\otimes\hdots\otimes E_{i_{12}}\varrho(0)E_{i_1}^\dagger\otimes\hdots\otimes E_{i_{12}}^\dagger.
\end{equation}
Again an optimisation over the product vectors $\ket{\Phi_S}$ introduced in Eq.(3) of the manuscript is performed in order to detect entanglement as reliably as possible with the hierarchy $\tau_{k,n}$ and the coefficients $a_i^{(k,n)}$ from Table~\ref{table}. In the present case we verified for $n\le8$ that
\be
|\Phi_S\rangle=(\ket{0}+i\ket{1})^{\otimes n}\otimes(\ket{0}-i\ket{1})^{\otimes n}/2^n
\ee
maximizes $\tau_{k,n}$ for all times independently of $k$ and $n$, and consequently used this choice of $\ket{\Phi_S}$ also for $n>8$. The decay properties depicted in Fig.~\ref{fig:decay} are thus known as analytic functions of $t$. %

Fig.~\ref{fig:decay} shows the decay of $\tau_{k,n}$ for $n=12$ and $k$ ranging from $2$ to $12$, where the substantially different behavior of $k$-partite entanglement for different $k$ is noticeable.
The instant $t_s$ at which $\tau_{k,n}$ becomes negative, so that no $k$-partite entanglement is identified anymore,
is depicted in the inset as a function of $k$.
As can be seen in this specific example $12$-partite entanglement can be observed until $\gamma t\simeq 0.12$ only, but the life-time of $10$-partite entanglement is already more than $2.5$ times longer. Furthermore, the lifetime of bipartite entanglement ($\gamma t\simeq 2.61$) exceeds that of genuine $12$-partite entanglement by a factor of $\approx$ 22, {\it i.e.} by more than an order of magnitude. The very different dynamical behavior that the various $k$-partite entanglement feature shows that few-partite entanglement can behave very differently from multi-partite entanglement.

\subsection{Identification of three-body interactions}\label{sec:slopes}

Finally, as an example of how detecting $k$-partite entanglement can provide insight on the dynamical features of a given system, let us demonstrate how the present hierarchies can be used to verify the existence of three-body interactions whose engineering is currently actively debated \cite{Mizel2004,Mazza2010}. 

Since typically more pairwise interactions than interactions among triples are required to create a $k$-partite entangled state, the onset of entanglement growth starting from a separable state can be expected to be a discriminator between these two types of interaction. For sufficiently short times entanglement grows in a monomial fashion $\tau_{k,n}(|\Psi(t)\rangle)\propto t^\alpha$ and, for a given system size $n$ and a suitably chosen $k$, the exponent $\alpha$ is a signature that discriminates two and three-body interactions, independently of their strength or specific form. This is shown here with the exemplary case of tripartite entanglement in a five-body system ($k$=3, $n$=5).

\begin{figure}
\includegraphics[width=0.4\textwidth]{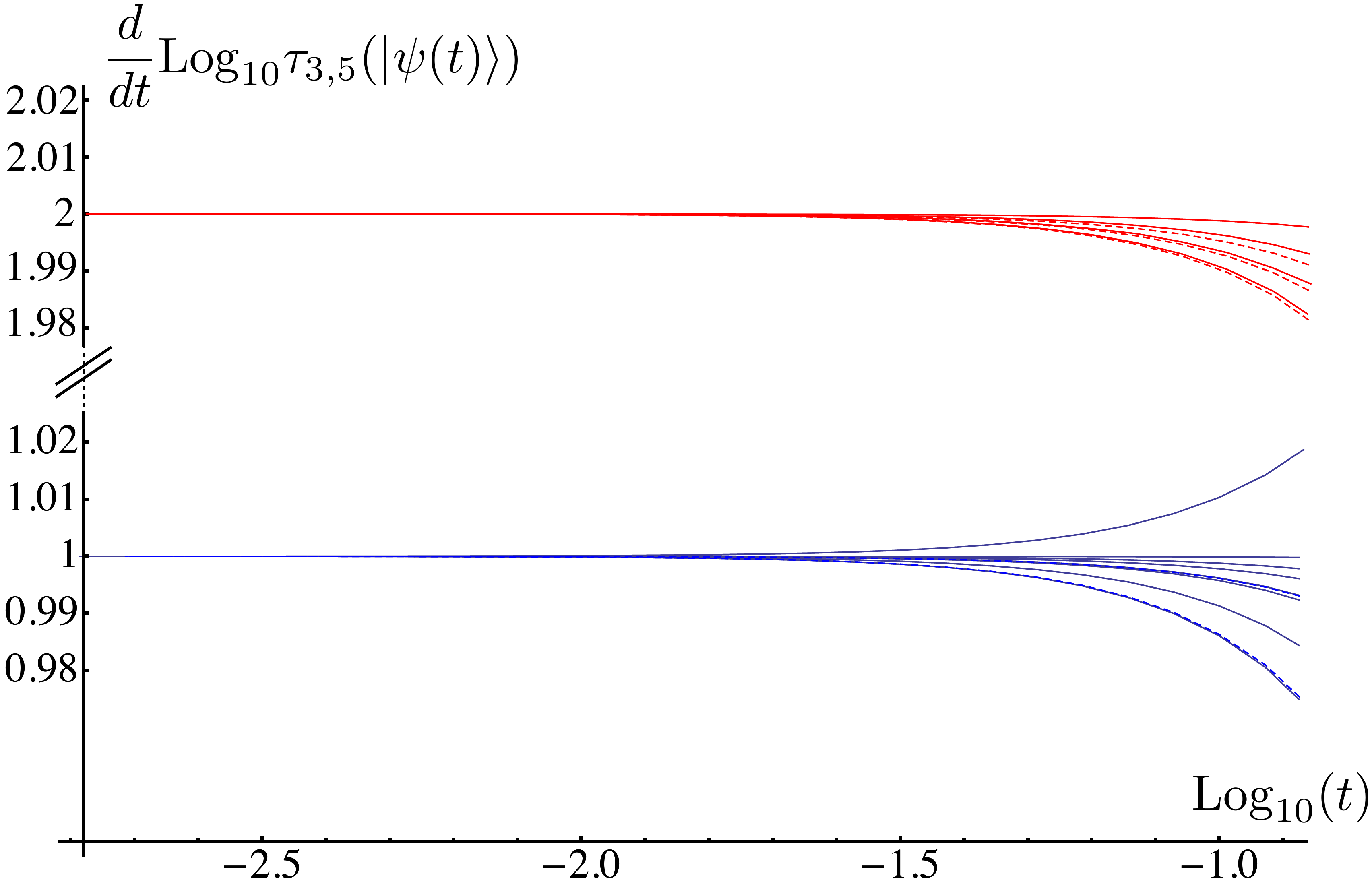}
\caption{\label{fig:slopes}
Comparison between the exponential growth of $\tau_{3,5}(|\psi(t)\rangle)$ in the case of a two-body (red) or three-body (blue) nearest neighbour spin-spin interactions as given in eq.(\ref{eq:randham}). Solid lines correspond to open boundary conditions, i.e. a chain; dashed lines to periodic boundary conditions, i.e. a ring. For short times the growth is monomial in time $\propto t^\alpha$ and the exponent is clearly shown to depend only on the type of the interaction, with all two-body interaction converging to $\alpha$=2 whereas $\alpha$=1 for three-body interactions, such that $\alpha$ allows to discriminate between them.}
\end{figure}

We consider the dynamics generated by Hamiltonians of the following form:

\begin{eqnarray}
\begin{aligned}
\label{eq:randham}
H_{2b}&=\sum_{k} \lambda_k \sigma_i^{(k)}\sigma_j^{(k+1)},\\
H_{3b}&=\sum_{k} \lambda_k  \sigma_i^{(k)}\sigma_j^{(k+1)}\sigma_\ell^{(k+2)}
\end{aligned}
\end{eqnarray}
where $\sigma_{i},\sigma_{j},\sigma_{\ell}$ are taken randomly from the three Pauli matrices and $\lambda_k$ are random coupling coefficients. A specific realization of these Hamiltonians will then generate the time evolution of an initially separable state $|\psi(0)\rangle=\ket{11111}$.

Fig.~\ref{fig:slopes} shows how indeed the growth of $\tau_{3,5}$ is monomial in time, as the time derivative of $\mathrm{Log}_{10}(\tau_{3,5})$ approaches a constant value for $t\rightarrow 0$. The curves corresponding to two-body interaction Hamiltonians ($H_{2b}$) are depicted in red, those originating from three-body interactions ($H_{3b}$) in blue: it is neatly shown how the exponents $\alpha$ converge to a value which depends exclusively on the nature of the interaction, independently of its details. For all two-body Hamiltonians $\alpha$ converges to the value $\alpha$=2, whereas for all the three-body interactions $\alpha$ converges to 1. Only for larger times signatures of the specific realisation become apparent, as deviations from the asymptotic values. This approach therefore provides a reliable method to witness whether a given interaction is of two or three-body nature. What is shown in Fig.~\ref{fig:slopes} is in sharp contrast to bipartite entanglement, which grows linearly in time both for two-body and three-body interactions and is thus of no help. This emphasises how different tasks may require probing $k$-partite entanglement for 2$\leq$$k$$\leq$$n$, and that the access to various $k$-partite entanglement allows us to grasp underlying dynamical features.

\bibliography{../../Bib/referenzen,../../Bib/libraryjabref}

\end{document}